\documentclass[letterpaper, 10 pt, conference]{ieeeconf}  

\IEEEoverridecommandlockouts                              
\makeatletter
\let\NAT@parse\undefined
\makeatother

\usepackage[dvipsnames]{xcolor}

\newcommand*\linkcolours{ForestGreen}

\usepackage{times}
\usepackage{graphicx}
\usepackage{amssymb}
\usepackage{gensymb}
\usepackage{amsmath}
\usepackage{breakurl}

\usepackage{url,hyperref}
\hypersetup{
colorlinks,
linkcolor=\linkcolours,
citecolor=\linkcolours,
filecolor=\linkcolours,
urlcolor=\linkcolours}

\usepackage{algorithm}
\usepackage{algorithmic}

\usepackage[labelfont={bf},font=small]{caption}
\usepackage[none]{hyphenat}

\usepackage{mathtools, cuted}

\usepackage[noadjust, nobreak]{cite}

\usepackage{tabularx}
\usepackage{amsmath}

\usepackage{float}

\usepackage{pifont}

\newcolumntype{Y}{>{\centering\arraybackslash}X}

\usepackage[]{placeins}

\usepackage{placeins}

\usepackage{tikz}

\usepackage[framemethod=tikz]{mdframed}

\usepackage{afterpage}

\usepackage{stfloats}

\usepackage{atbegshi}
\newcommand{\handlethispage}{}
\newcommand{\discardpagesfromhere}{\let\handlethispage\AtBeginShipoutDiscard}
\newcommand{\keeppagesfromhere}{\let\handlethispage\relax}
\AtBeginShipout{\handlethispage}

\usepackage{comment}

\title{\LARGE \bf
A remark on a paper of Krotov and Hopfield
}

\author{Fei Tang$^{1}$ and Michael Kopp$^{2}$
\thanks{$^{1}$HERE Technologies, Z\"urich,
Email: fei.tang@here.com}%
\thanks{$^{2}$IARAI, Vienna,
Email: michael.kopp@iarai.ac.at}%
}

\begin{document}

\maketitle
\thispagestyle{empty}
\pagestyle{empty}

\noindent
\begin{abstract}
\noindent
In their recent paper titled ``Large Associative Memory Problem in Neurobiology and Machine Learning" \cite{krotov2021large} the authors gave a biologically plausible microscopic theory from which one can recover many dense associative memory models discussed in the literature. We show that the layers of the recent "MLP-mixer" \cite{tolstikhin2021mlpmixer} as well as the essentially equivalent model in \cite{melaskyriazi2021need} are amongst them.
\end{abstract}

\section{INTRODUCTION}
\noindent
Krotov et al~\cite{krotov2021large} proposed a continuous network of feature and memory neurons with symmetric synaptic connections between them. The time evolution of the neurons are described by the following dynamic equations.
\begin{align}
    \tau_f \frac{dv_i}{dt} &= \sum_{\mu = 1}^{N_h}\xi_{i\mu}f_{\mu} - v_i + I_i  \nonumber\\
\tau_h \frac{dh_{\mu}}{dt} &= \sum_{i=1}^{N_f}\xi_{\mu i}g_i - h_{\mu}
\end{align}
An energy function $E(t)$ was stipulated:
\begin{align}
[\sum_{i = 1}^{N_f}(v_i - I_i)g_i - L_v] + [\sum_{\mu=1}^{N_h}h_{\mu}f_{\mu} - L_h] - \sum_{\mu i} f_{\mu}\xi_{\mu i}g_i
\end{align}
They showed that the Dense Associative Memory model~\cite{krotov2016dense} and Modern Hopfield Networks model~\cite{ramsauer2021hopfield} can be recovered from this theory using particular Lagrangian functions $L_v$ and $L_h$ as Model A and B, respectively. A Model C was also discussed as having spherical normalization in the feature layer, but without analogue models in the literature. We show in this note that the newly published "MLP-mixer" \cite{tolstikhin2021mlpmixer} can be partially recovered using essentially the Lagrangian for Model C.
\section{DERIVING THE MLP MIXER LAYER}
\noindent
We propose the following slightly amended Lagrangian $L_v$ to Model C in \cite{krotov2021large}
\begin{align}
    L_v &= \sqrt{\sum_i (v_i - \bar v)^2}\qquad
    \text{where }\bar v = \frac{\sum_i v_i}{N_f}
\end{align} and the same $L_h$, i.e. $L_h = \sum_{\mu}F(h_{\mu})$. This implies
\begin{align}
    f_{\mu} &= F'(h_{\mu})\nonumber = f(h_\mu)\\
    g_i &= \frac{\partial L_v}{\partial v_i} = \frac{v_i - \bar v}{\sqrt{\sum_l (v_l-\bar v)^2}}
\end{align}
Equation 20 in \cite{krotov2021large} still holds for this choice of $L_v$, as
\begin{align}\label{modelc_zeromode}
\sum_j M_{ij}v_j &= \frac{1}{\sqrt{\sum_l(v_l - \bar v)^2}}\sum_j (\delta_{ij} - \frac{1}{N})v_j\nonumber\\
&- \frac{v_i - \bar v}{(\sqrt{\sum_l(v_l - \bar v)^2})^3}\sum_j(v_j - \bar v) v_j\nonumber\\
&=0
\end{align}
Thus the dynamic equation for feature neurons (equation 22 in \cite{krotov2021large}) with arbitrary $\alpha$ still holds:
\begin{align}\label{modelc_updateeqn}
\tau_f\frac{dv_i}{dt} = \sum_{\mu}\xi_{i\mu}f(\sum_j \xi_{\mu j}g_i) - \alpha v_i
\end{align}
If we choose $\alpha=0$, $dt=\tau_f$ and plug in the $g_i$ the update equation becomes:
\begin{align}\label{updateeqn_mixlayer}
    v_i^{t+1} = v_i^{t} + \sum_{\mu}\xi_{i\mu}f\left(\sum_j \xi_{\mu j} \frac{v_j^t - \bar v^t}{\sqrt{\sum_k (v_k^t - \bar v^t)^2}}\right)
\end{align}
This corresponds to each of the Mixer layers in \cite{tolstikhin2021mlpmixer}
\begin{align}
    U_{*, i} &= X_{*, i} + W_2\sigma(W_1 \text{LayerNorm}(X)_{*, i}, \text{for}\, i = 1\dots C,\nonumber\\
    Y_{j, *} &= U_{j, *} + W_4\sigma(W_3 \text{LayerNorm}(U)_{j, *}, \text{for}\, j = 1\dots S
\end{align}
where $\frac{v_j^t - \bar v^t}{\sqrt{\sum_k (v_k^t - \bar v^t)^2}}$ corresponds to the LayerNorm(X), $\sum_j \xi_{\mu j}\cdots$ corresponds to $W_1$ or $W_3$, $\sigma = f=F'$, $\sum_{\mu}\xi_{i\mu}\cdots$ corresponds to $W_2$ or $W_4$ and $v_i^{t} + \cdots$ corresponds to the skip connection.

\section{DISCUSSION}
\noindent
In \cite{krotov2021large}, the matrix $\xi_{\mu i}$ is assumed to be symmetric. Thus the above correspondence places a strict limit on the Mixer layer in that $W_1$ and $W_2$ are transposed to each other, and same for $W_3$ and $W_4$. It is possible to drop the symmetric assumption for the special case of Model C which makes an exact one-to-one mapping possible. We leave these derivations to future work.

\bibliographystyle{ieeetr}
\bibliography{remark}

\begin{thebibliography}{1}

\bibitem{krotov2021large}
D.~Krotov and J.~Hopfield, ``Large associative memory problem in neurobiology
  and machine learning,'' 2021.

\bibitem{tolstikhin2021mlpmixer}
I.~Tolstikhin, N.~Houlsby, A.~Kolesnikov, L.~Beyer, X.~Zhai, T.~Unterthiner,
  J.~Yung, A.~Steiner, D.~Keysers, J.~Uszkoreit, M.~Lucic, and A.~Dosovitskiy,
  ``Mlp-mixer: An all-mlp architecture for vision,'' 2021.

\bibitem{melaskyriazi2021need}
L.~Melas-Kyriazi, ``Do you even need attention? a stack of feed-forward layers
  does surprisingly well on imagenet,'' 2021.

\bibitem{krotov2016dense}
D.~Krotov and J.~J. Hopfield, ``Dense associative memory for pattern
  recognition,'' 2016.

\bibitem{ramsauer2021hopfield}
H.~Ramsauer, B.~Schäfl, J.~Lehner, P.~Seidl, M.~Widrich, T.~Adler, L.~Gruber,
  M.~Holzleitner, M.~Pavlović, G.~K. Sandve, V.~Greiff, D.~Kreil, M.~Kopp,
  G.~Klambauer, J.~Brandstetter, and S.~Hochreiter, ``Hopfield networks is all
  you need,'' 2021.

\end{thebibliography}


\end{document}